\documentclass[pre, preprint, showpacs, showkeys]{revtex4-1}
\usepackage{amsmath}
\usepackage{amssymb}
\usepackage{amsfonts}
\usepackage[dvips]{graphicx}
\usepackage{epsfig}
\usepackage[]{subfigure}
\usepackage[dvips]{color}
\usepackage{color}
\usepackage{graphicx}
\usepackage{latexsym}
\usepackage{amscd}
\usepackage{times}
\usepackage{natbib}

\def\kon{K_\textrm{on}}
\def\koff{K_\textrm{off}}

\def\la{\langle}
\def\ra{\rangle}

\newcommand{\beq}{\begin{equation}}
\newcommand{\eeq}{\end{equation}}

\definecolor{mypink1}{rgb}{0.858, 0.188, 0.478}
\definecolor{mypink2}{RGB}{219, 48, 122}
\definecolor{mypink3}{cmyk}{0, 0.7808, 0.4429, 0.1412}
\definecolor{mygray}{gray}{0.6}
\definecolor{yellow}{cmyk}{0.0, 0.0, 1.0, 0.0}

\definecolor{mycolor}{rgb}{0.9, 0.5, 0.2}
\definecolor{yellow}{cmyk}{0.0, 0.0, 1.0, 0.0}

\newcommand{\fat}{\sffamily\textbf}

\usepackage{soul}
\setstcolor{blue}

\begin{document}

\title{
Anomalous diffusion and FRAP dynamics in the random comb model
}

\author{S. B. Yuste,$^{1}$ E. Abad,$^{2}$ and A. Baumgaertner$^{1}$}
\affiliation{
  $^{1}$ Departamento de F\'{\i}sica and Instituto de Computaci\'on Cient\'{\i}fica Avanzada (ICCAEX) \\
 Universidad de Extremadura, E-06071 Badajoz, Spain \\
$^{2}$ Departamento de F\'{\i}sica Aplicada and Instituto de Computaci\'on Cient\'{\i}fica Avanzada (ICCAEX) \\ Centro Universitario de M\'erida \\ Universidad de Extremadura, E-06800 M\'erida, Spain
}

\begin{abstract}
We address the problem of diffusion on a comb whose teeth display a varying length. Specifically, the length $\ell$ of each tooth is drawn from a probability distribution displaying the large-$\ell$ behavior $P(\ell) \sim \ell^{-(1+\alpha)}$ ($\alpha>0$). Our method is based on the mean-field description provided by the well-tested CTRW approach for the random comb model, and the obtained analytical result for the diffusion coefficient is confirmed by numerical simulations. We subsequently incorporate retardation effects arising from binding/unbinding kinetics into our model and obtain a scaling law characterizing the corresponding change in the diffusion coefficient. Finally, our results for the diffusion coefficient are used as an input to compute concentration recovery curves mimicking FRAP experiments in comb-like geometries such as spiny dendrites. We show that such curves cannot be fitted perfectly by a model based on scaled Brownian motion, i.e., a standard diffusion equation with a time-dependent diffusion coefficient. However, differences between the exact curves and such fits are small, thereby providing justification for the practical use of models relying on scaled Brownian motion as a fitting procedure for recovery curves arising from particle diffusion in comb-like systems.

\end{abstract}

\date{\today}


\maketitle

\section{Introduction}

Random walks of particles in complex environments play a central role as models for anomalous transport processes in physics, biology and chemistry.  In this context, a wealth of experimental evidence shows that slowing-down of particle diffusion is a common occurrence \cite{KlagesBook}. In order to set up a random walk model  tailored for the experimental situation at hand,  one would ideally like to have a detailed knowledge of the microscopic mechanisms underlying the slowed-down diffusion.  However, this remains a challenging issue, since typically a number of possible factors responsible for the onset of subdiffusive regimes coexist and it is often difficult to identify the dominant effect(s).  Among such factors are strong geometric constraints associated with fractal, labyrinthine an disordered environments, viscoelastic effects, excluded volume interactions due to obstacles, crowding effects,  binding/unbinding processes of different nature, cage effects due to electrostatic interactions, etc.

In order to capture the phenomenology leading to subdiffusion, three types of models are often invoked, namely \cite{Sokolov2012}: random walks in complex geometries, random walks with non-independent increments (resulting in anti-persistence effects), and walks displaying  memory effects (aging). Each model class differs in its statistical properties from the other two, yet there may be instances in which different models yield a fairly similar behavior of a specific quantity. This fuels the debate as to which model class is the most appropiate one to account for the behavior observed in a given experiment, and the use of hybrid models in this context is not uncommon \cite{Meroz2015}. However, even if one chooses to work with a model belonging to a single class, one still has to deal with many challenges. Focusing on the category of walks in complex geometries, attempts to shed light on the relationship between transport properties and the topological details of the embedding support often face considerable difficulties. For example, complex geometries often lead to non-trivial behavior, such as the onset of crossover regimes between normal diffusion and anomalous diffusion.  In the particular case of branching geometries, such effects are observed because of the large time needed by the diffusing walkers to explore the complexity of the environment in full detail.

In the above context, comb and comb-like models \cite{White1984, Goldhirsch1986, Weiss86, Havlin87, Haus1987, Bouchaud90, AvrahamBook2005} use simple, idealized geometries to capture the essential features of transport in natural branching structures and, more generally, to mimic transport properties of disordered networks.  The basic idea is to distribute a number of vertical teeth along a one-dimensional line (the backbone), and to allow random walkers diffuse throughout the resulting structure, whereby occasional excursions along the teeth may be viewed as trapping events which slow down the particles' motion along the backbone. The simplest situation corresponds to the regular comb model \cite{Weiss86}, where both the separation between adjacent teeth and the length of each tooth are constant quantities. In the general case, the separation and the tooth length are random variables.

A comprehensive list of examples for which comb and comb-like structures are relevant
can be found in Refs.~\cite{Mendez2015} and \cite{Berezhkovskii2015}, including spiny dendrites, diffusion of drugs in the circulatory system, energy transfer in polymer systems, etc. Other examples include oxygen exchange in lungs and water circulation in river networks \cite{Fleury2001}. While the regular comb model was originally devised to study anomalous transport properties in percolation clusters, more sophisticated extensions thereof \cite{Arkhincheev1999, Arkhincheev2000, Arkhincheev2002a, Iomin2005, Arkhincheev2007a, Arkhincheev2007,  Zaburdaev2008, Rebenshtok2013, Mendez2013, Sandev2015, Sandev2015a, Mendez2015, Berezhkovskii2015, Illien2016} have been used to make the phenomenology richer and to account for the presence of heterogeneities and spatial disorder in real systems.

In the present work, we shall focus on the problem of diffusion on a comb displaying a random variability in the lengths of its teeth, a subject that has already been discussed to some extent in previous works  \cite{Havlin87, Havlin87b, Aslangul1994, Pottier1994a, Pottier1995, Balakrishnan1995, Arkhincheev2002a, AvrahamBook2005, Durhuus2006, Elliott07}.  Specifically, the length $\ell$ of each tooth is drawn from a distribution whose large-$\ell$ behavior is given by the asymptotic form $P(\ell) \sim \ell^{-(1+\alpha)}$. The characteristic exponent $\alpha>0$ can be used to control the rate at which particles diffuse along the backbone. On the other hand, the random comb model can be regarded as a somewhat raw picture of real-world comb-like systems such as spiny dendrites, which indeed show a variability in the spine length. At least in some cases, the latter appears to follow a power-law distribution. For example, Fig.~4 in Ref.~\cite{Ruszczycki2012} shows that the spine length follows a non-Gaussian distribution which turns out to be well fitted by a power law in the appropriate regime (data not shown).

In an extended version of our model, we shall also consider the effect of combining particle transport with binding/unbinding events. In biological comb-like systems such as spiny dendrites in Purkinje neurons, the mobility of signaling species such as calcium ions is strongly hindered by morphological factors leading to signal compartmentalization in single spines \cite{Schmidt2003} (corresponding to comb teeth in the simplified picture of our model).  However, the range of action of free calcium ions is also severely limited by the effect of binding proteins. This exemplifies the relevance of retardation effects associated with binding/unbinding kinetics in biological systems.

A popular experimental technique for the characterization of diffusive transport concomitant with binding/unbinding kinetics is based on so-called FRAP (Fluorescence Recovery After Photobleaching) experiments. In these experiments, the diffusing molecules in the system are first fluorescently tagged, and then those molecules found in a small region (``the bleached spot'') are photobleached by a brief, intense laser pulse. The resulting relaxational dynamics leads to a refilling of the bleached spot and to the recovery of the associated fluorescent signal, which is monitored with the help of suitable microscopy techniques \cite{McNally05, McNally10}.

Recently, the experimental characterization of FRAP kinetics in comb-like systems such as spiny dendrites in neurons has attracted some interest \cite{Zheng2011, Schmidt2003, Schmidt2007}. Typically, individual spines are photobleached and the subsequent concentration recovery is modeled by means of one-dimensional effective transport equations. Here, we shall consider a more general setting in which the propagation of unbleached particles takes place throughout the entire comb geometry, including both the teeth and the backbone.

Despite the intensive analytical and computational work performed on FRAP models so far \cite{Saxton01, McNally04, McNally08, Lubelski2008, McNally10, Yuste2014}, the theoretical characterization of FRAP dynamics on comb-like structures does not seem to have been dealt with.  In the present work, we shall address this issue in detail and obtain analytic and numerical results for concentration recovery curves. As it turns out, these recovery curves cannot be reproduced \emph{exactly} by means of a standard diffusion equation with a suitably chosen time-dependent diffusion coefficient. However, the resulting discrepancy appears to be small, suggesting that such models may be acceptable for certain purposes. 

A first step towards the solution of the FRAP problem on the random comb is the calculation of the diffusion coefficient of the particles. An early work by Havlin et al. \cite{Havlin87b} showed that in the random comb model a crossover from subdiffusion to normal diffusion takes place when the decay exponent $\alpha$ of the power law exceeds the threshold value $\alpha=1$.  For $0<\alpha<1$ this random comb model was shown to yield anomalous diffusion with a characteristic exponent $\gamma=(1+\alpha)/2$.

\begin{figure} [thb]
  \begin{center}
\includegraphics[width=0.4\textwidth, angle=0]
{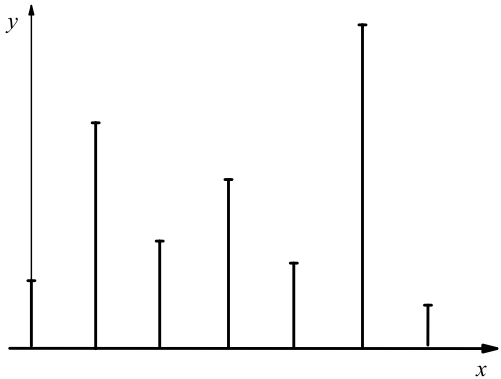}
\caption{
 Scheme of a random comb structure with equally spaced teeth of varying length.}
\label{CombFig}
  \end{center}
\end{figure}

Other works focusing on the behavior of the diffusion exponent in the random comb model are also found in the literature \cite{Balakrishnan1995,Arkhincheev2002a}. However, the behavior of the \emph{diffusion coefficient} for the particle motion along the backbone of the comb is only partially known. For the case of normal diffusion ($\alpha>1$), the diffusion coefficient was computed via different methods \cite{Revathi1993, Balakrishnan1995, Revathi1996}. In the subdiffusive regime ($\alpha<1$), the diffusion coefficient of the \emph{subset} of particles located on the backbone is known \cite{Aslangul1994,Lubashevskii1998}. This quantity can be formalized as follows. Let us take the backbone as the $x$-axis, and let the $y$-axis denote the vertical direction along which the comb teeth extend (see Fig. ~\ref{CombFig}). The  mean square displacement (MSD) of the subset of particles on the backbone can then be expressed as $\langle x^2\rangle =\int x^2 {\cal G}(x,y=0,t) \, dx /\int  {\cal G}(x,y=0,t) \, dx$, where ${\cal G}(x,y,t)$ is the probability density (Green's function) to find a walker which started from a given initial condition at position $(x,y)$ at time $t$. In contrast, the case of interest in the context of the FRAP relaxation problem addressed in Sec.~VI refers to the diffusion properties of the full set of particles, i.e., those found on the backbone \emph{and} on the teeth. In terms of the Green's function, the corresponding MSD of a walker is $\langle x^2\rangle = \int\int x^2 {\cal G}(x,y,t)\, dx dy$.  This case was studied in Ref.~\cite{Balakrishnan1995}, but the expression for the associated diffusion coefficient was only given for the normal diffusion case ($\alpha>1$). 
%
%
%

The present work aims to fill this gap by providing an explicit expression for the diffusion coefficient in various regimes and subsequently validating it by means of extensive Monte Carlo simulations. To this end, we shall resort to the well-tested mean-field approach provided by the so-called continuous time random walk (CTRW) model \cite{Weiss86,Havlin87b,Balakrishnan1995,Lubashevskii1998,Arkhincheev2002a}. In this approach,  particle excursions along the teeth are considered to be ``a waste of time'' as far as diffusion along the backbone is concerned. Since a particle moving randomly along a tooth does not experience any change in its horizontal position,  the waiting time between subsequent changes in the $x$-coordinate will follow a distribution which is directly related to the tooth length distribution $P(\ell)$.

The remainder of the present work is organized as follows:  Sec.~II gives a detailed definition of the random comb model and shows how it is implemented in our numerical simulations. Sec.~III presents numerical results for the time-dependence of the MSD along the backbone of the comb. Sec.~IV deals with the CTRW-based method used for the analytical computation of the diffusion coefficient. Sec.~V discusses the role of retardation effects arising from the combination of transport and binding/unbinding events. Sec.~VI is devoted to a comprehensive analytical and numerical treatment of FRAP dynamics in the random comb model. The results in this section are based on the well-known asymptotic equivalence between the CTRW model and the fractional diffusion equation \cite{Metzler00}. Finally, Sec.~VII gives a summary of the main conclusions and outlines some avenues for future research.  Technical details concerning the calculation of the waiting time probability density function (pdf) associated with the most general form of the diffusion coefficient are given in Appendix A. The solution of the boundary value problem for FRAP dynamics by means of the Green's function formalism is given in Appendix B. 

\section{Definition of the model and simulation procedure}

In order to address the problem of diffusion on the random comb, we first introduce the regular comb model. As already mentioned, the regular comb consists of a backbone and equally spaced teeth of a fixed length. A particle diffusing along the backbone may encounter a tooth and perform an excursion along it before eventually returning to the backbone. The case of a regular comb with infinite tooth length $\ell \to \infty$ was discussed in Ref.~\cite{Weiss86}.  In the appropriate regime, this model yields subdiffusive behavior with anomalous diffusion exponent equal to 1/2, that is, $\langle x^2 \rangle \propto t^{1/2}$.

If the regular spacing between the teeth is kept (or, more generally, if this spacing follows a probability distribution with finite variance) yet random changes in the tooth length are allowed (see Fig.~\ref{CombFig}), one obtains a specific class of random comb models displaying a surprisingly rich phenomenology. In particular, the length $\ell$ may be chosen independently for each tooth by drawing its value from a distribution whose asymptotic behavior is $P(\ell) \sim \ell^{-(1+\alpha)}$, where $\alpha>0$. In what follows, we focus on this specific case, which was already discussed in Refs.~\cite{Havlin87} and \cite{Havlin87b}. In those references it was shown that the system exhibits anomalous subdiffusion along the backbone axis for $0<\alpha <1$, i.e., one has $\langle x^2(t)\rangle \sim t^{\gamma}$ with $\gamma = (1+\alpha)/2$. In contrast, for $\alpha>1$ there is a crossover to normal diffusion, that is, $\langle x^2(t)\rangle \sim t$. At the crossover value $\alpha=1$ there is a logarithmic correction, and hence $\langle x^2(t)\rangle \sim t/\ln(t)$. Thus, the decay exponent $\alpha$ of the tooth length distribution can be used to tune the value of the diffusion exponent in the range $1/2\le \gamma \le 1$.

Our first goal will be to discuss the results of extensive Monte Carlo simulations implemented on the random comb structure depicted in Fig.~\ref{CombFig}. In order to carry out the simulations, we discretized the comb geometry as follows. The unit length (lattice spacing) was chosen to be equal to the distance between two consecutive teeth, and each tooth consisted of a randomly chosen integer number of lattice spacings.

The choice of the discretized tooth length was implemented as follows.  We attached a tooth of integer length $\ell_k=\left \lfloor r^{-\alpha} \right \rfloor$ to each backbone site $k$,
where $r$ denotes a uniformly distributed random number ($0 < r < 1$) and  $\left \lfloor \xi \right \rfloor= \max\, \{m\in\mathbb{Z}\mid m\le \xi\}$ stands for the floor function. Thus, the probability $P(\ell)$ that a randomly chosen tooth had a length of  exactly $\ell$ lattice spacings is
\begin{equation}
\label{PLdiscreta}
\mathcal{P}(\ell)=\ell^{-\alpha}-(\ell+1)^{-\alpha}.
\end{equation}
As a result of the above prescription, the tooth length follows approximately the pdf
\beq
P(\ell) = \alpha \,\ell^{-(1+\alpha)}.
\label{Pl}
\eeq
This expression becomes increasingly accurate as $\ell$ becomes larger.

A collection of random walkers were then allowed to perform nearest-neighbor jumps on the discretized system at regular time intervals (the time unit was taken to be the fixed waiting time between two consecutive jumps).  Specifically, the walk of each particle on the random comb was implemented as follows. When a given walker was on a tooth ($y> 0$), its motion was restricted to the vertical direction ($x(t) = const$). As soon as the walker returned to the backbone ($y = 0$), it could either jump back to $y=1$ with probability 1/2, or move along the $x$-axis with probability 1/2 (to the left with probability 1/4 or to the right with probability 1/4).

The boundary conditions in $x$- and $y$-directions were implemented as follows. Since each realization of the comb geometry could contain a finite number of teeth only, we introduced periodic boundary conditions along the $x$-direction in order to preserve the translational invariance of the system. We thus considered a finite system of  $N$ random walkers in a  ``periodic box'' of $M$ length units, where typically $200 \leq M \leq 1600$ and $1000 \leq N \leq 8000$. For sufficiently large values of $M$, typical diffusion properties of individual walkers no longer display a significant size dependence if the simulation time is not too long, thereby ensuring that the typical diffusion distance is small with respect to the linear system size. Thus, the behavior of the finite system is expected to become indistinguishable from that of the corresponding infinite system.

Particle jumps in $y$ direction were limited by the finite tooth length $\ell$. Whenever a given particle would reach the end of a tooth ($y(t)=\ell$), at the next time step $t+\Delta t$,  the particle would either moved back to site $\ell-1$, or else attempt to perform a jump beyond the end of the tooth $y(t+\Delta t)= \ell+1$. In the latter case, it was ``reflected back'', as a result of which it remained in the same position [$y(t+\Delta t) = y(t)=\ell$].

In order to speed up the simulations, several walkers were randomly scattered along the backbone and then launched simultaneously. In this case, the Monte Carlo time step was defined as $\Delta t = 1/N$, where $N$ is the number of walkers \cite{BinderBook}. At each time step, one particle was randomly chosen and performed a jump (unless it attempted to ``exit'' a tooth, in which case it remained at its upper end). The selection of the particle was either sequential or random, both choices leading to similar results in the long time limit. Thus, a time unit corresponded to $N$ time steps, that is, to $N$ attempted movements of the walkers. With the above choice, the time unit $N \Delta t =1$ does not depend on the number of particles. Typical simulation times were $t\le 10^8$, whereas maximum excursions along the $y$-axis were of the order of 400 lattice spacings.

\section{Onset of anomalous diffusion: numerical study of the long-time asymptotics and transient regimes }

In order to study diffusive transport along the backbone, we computed the MSD at time $t$ by generating $n_q$ independent realizations of the comb geometry and then letting $N$ non-interacting walkers simultaneously evolve in each of them. The MSD is given by the formula 
\beq
\la x^2(t) \ra = \frac{1}{n_q} \sum_{s=1}^{n_q}\frac{1}{N} \sum_{j=1}^N \lbrack x_j^{(s)}(t) - x_j^{(s)}(0)\rbrack^2,
\label{doubleavMSD}
\eeq
where $x_j^{(s)}(t)$ denotes the $x$-coordinate of the $j$-th walker diffusing in the $s$-th
realization of the quenched disorder. Unless otherwise specified, it is understood that all the walkers were placed at random on the backbone at the beginning of each run. For specific calculations, the average over quenched disorder was typically performed over $n_q = 50$ configurations, corresponding to 50 different landscapes $\{\ell_k\}$.  

\begin{figure} [thb]
  \begin{center}
\includegraphics[width=0.5\textwidth, angle=0]
{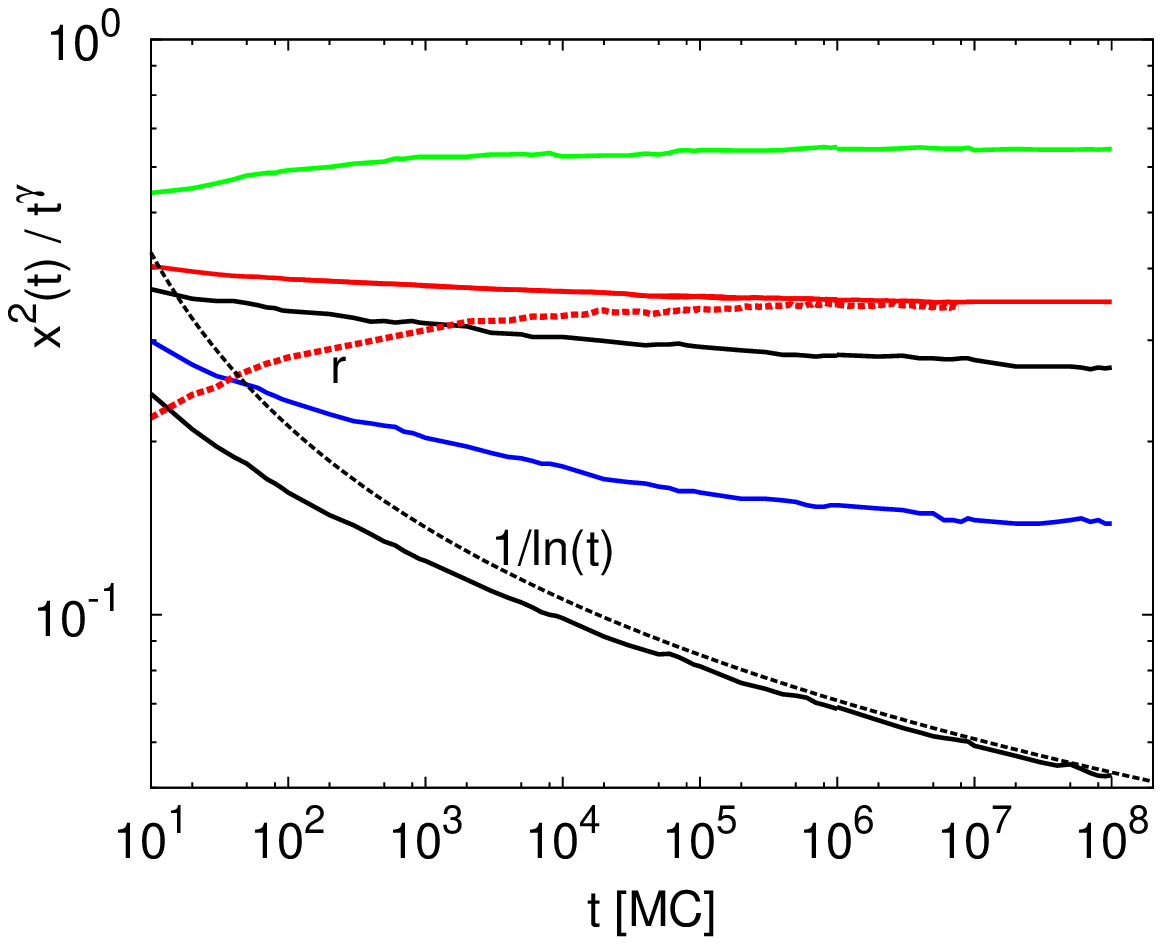}

\includegraphics[width=0.5\textwidth, angle=0]
{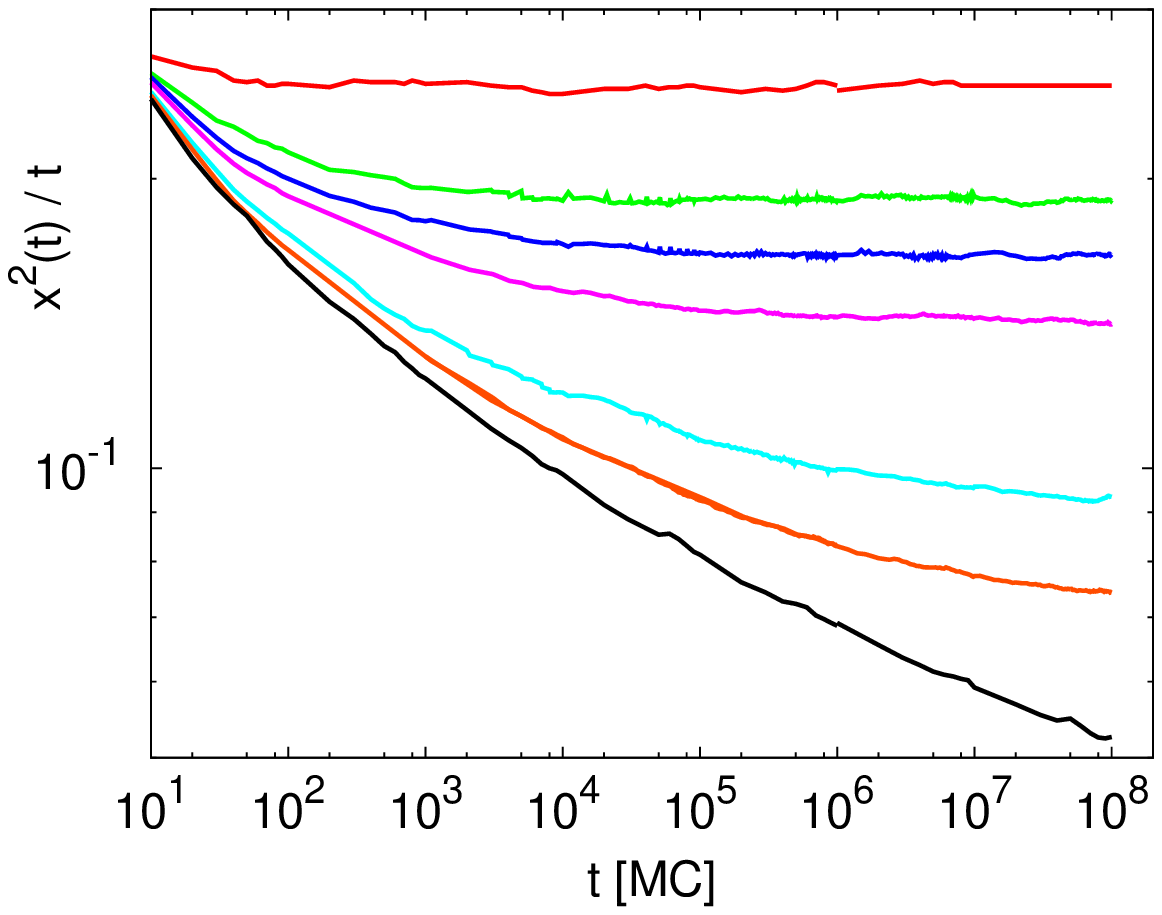}

\vspace*{-0.2cm}
\caption{\footnotesize
\baselineskip 0.3cm
Double logarithmic plots for the time evolution of $\langle x^2 \rangle/t^\gamma$ as obtained from numerical simulations. The anomalous diffusion exponent is assumed to be given by the theoretical prediction, i.e., $\gamma = (1+\alpha)/2$ for $\alpha<1$ and $\gamma=1$ for $\alpha\ge 1$. All particles are initially located on the $x$-axis ($y(0)=0$). The different curves correspond, from top to bottom to $\alpha=0.2, 0.5, 0.6, 0.8 \mbox{ and } 1$ (top figure) and to $\alpha=99.0, 2.0, 1.7, 1.5, 1.2, 1.1. \mbox{ and } 1$ (bottom figure). The additional dashed curve corresponding to the behavior of $1/\ln(t)$ is seen to match the asymptotic long-time behavior when $\alpha=1$. For $\alpha=0.5$, the data represented by the curve denoted by ''r'' correspond to the case where the particles are initially distributed at random along the teeth, and no particles are placed on backbone sites.}
\label{Xt}
  \end{center}
\end{figure}

According to previous references \cite{Havlin87,Havlin87b}, the long-time behavior of the MSD is
\beq
\la x^2(t) \ra = D_0(\alpha)~t^{\gamma},
\label{x2t}
\eeq
with $\gamma=(\alpha+1)/2$ for $0<\alpha<1$ and $\gamma=1$ for $\alpha>1$. The above analytical prediction is corroborated by the results displayed in Fig.~\ref{Xt}, where the behavior of $\langle x^2 \rangle/t^\gamma$ is plotted for different values of $\alpha$.  In the long-time regime this quantity typically reaches a well-defined plateau. In contrast, no plateau is observed when $\alpha=1$ (this is precisely the $\alpha$-value at which a transition between anomalous diffusion and normal diffusion is observed). In this case, the quantity $\langle x^2 \rangle/t^\gamma\equiv\langle x^2 \rangle/t$ follows an inverse logarithmic law (see caption of Fig.~\ref{Xt}).

\begin{figure} [thb]
\begin{center}
\includegraphics[width=0.5\textwidth, angle=0]
{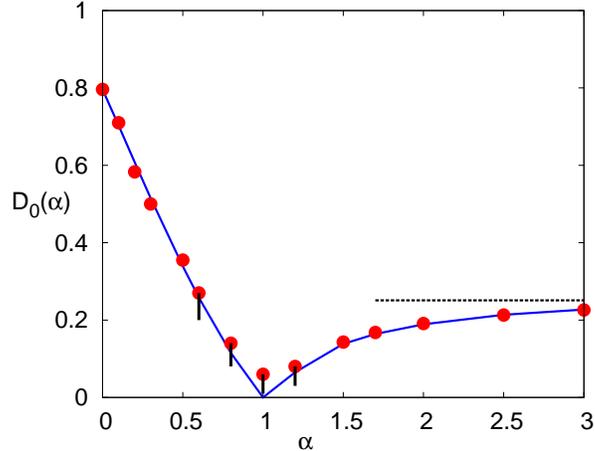}
\vspace*{-0.2cm}
\caption{\footnotesize
\baselineskip 0.3cm
Anomalous diffusion coefficient $D_0(\alpha)$ as a function of  $\alpha$. The dots correspond to simulation results. Those dots marked with vertical bars correspond to $\alpha$-values for which the simulation time is not sufficient for $\langle x^2 \rangle /t^\gamma$ to reach a plateau.  In such cases, the dots provide upper bounds for $D_0(\alpha)$. The solid curve corresponds to the theoretical expression given by  Eq.~\eqref{D0alpha4} for $0<\alpha<1$, and by Eq~\eqref{D0alphagt1} for $\alpha>1$. The dashed horizontal line denotes the asymptotic value $D_0(\alpha \to \infty)=1/4$, corresponding to a uniform teeth with $\ell=1$ everywhere. 
\label{D0}
}
\end{center}
\end{figure}

In those cases where the simulation time is long enough to observe a plateau, the asymptotic values of $\langle x^2 \rangle/t^\gamma$ obtained from the simulation data displayed in Fig.~\ref{Xt} can be used to estimate the values of the $\alpha$-dependent effective diffusion coefficient $D_0(\alpha)\equiv \mbox{lim}_{t\to\infty} \langle x^2 \rangle/t^\gamma $. In the approximate range $0.5 < \alpha < 1.5$,  the simulation time is not long enough to allow the system to reach a plateau. However, since $\langle x^2(t)\rangle/ t^{\gamma}$ decreases monotonically in time, the smallest value of this quantity can be used as an upper bound for $D_0$. The behavior of $D_0$ is shown in Fig.~\ref{D0}. The diffusion coefficient is seen to display non-monotonic behavior, first it decreases and then it increases with increasing $\alpha$.

We close this section with a comment on the role of the initial condition. According to our simulation results, the long-time value of $\langle x^2 \rangle/t^\gamma$ and the corresponding exponent $\gamma$ are not influenced by the specific initial condition. As an example, one may consider the case where all the particles are initially scattered at random along the teeth only, and no particles lie on the backbone (i.e., $y(0)>0$ for all the particles). The dashed curve denoted by ``r'' in Fig.~\ref{Xt} corresponds to one such initial condition implemented for the case $\alpha =0.5$. As expected, the MSD $\la x^2(t) \ra$ at short times is smaller than for our previous initial condition with all the particles lying on the backbone (see the solid curve displayed  in Fig.~\ref{Xt} for $\alpha=0.5$). The reason is of course that particles on the teeth must first return to the backbone in order to be able to contribute to the increase of $\la x^2(t) \ra$.

\section{Evaluation of the anomalous diffusion coefficient via the CTRW model}
\label{secDifuCTRW}

\subsection{General formalism}
As already mentioned, the values of the diffusion coefficient $D_0(\alpha)$ can be read off the plateaus of Fig.~\ref{Xt} for different values of $\alpha$. We now proceed to compute  $D_0(\alpha)$  analytically by means of the mean-field CTRW approximation for the random comb. Admittedly, the comb model displays quenched configurational disorder, implying that the tooth length distribution does not change in the course of the random walk. In contrast, the CTRW model can be regarded as an annealed version of the comb model in which the length of a given tooth is drawn anew from the corresponding distribution upon each revisitation of the walker. In line with a number of previous references (e.g. \cite{Aslangul1994,Balakrishnan1995}), we shall hereafter assume that the difference between the quenched system and the annealed system underlying the CTRW approach can be ignored as far as the leading asymptotic behavior of the diffusive process is concerned. As we shall see, this hypothesis will \emph{a posteriori} find strong support in the agreement between the analytical results obtained in the present section and the simulations results displayed in Sec.~III. 

As far as diffusion along the $x$-axis is concerned, the time spent by a walker traveling along the $y$-axis can be regarded as a waiting time between two consecutive steps along the backbone.  Thus, the movement of the walker along the $x$-axis can be described by means of the CTRW model. In this model, the waiting time distribution function $\psi(t)$ is the key quantity. Below, we show how to evaluate $\psi(t)$  for the random comb model.

Our method follows closely the one laid out in Ref.~\cite{Havlin87b} by  Havlin \emph{et al.}, which is based on the computation of the exact long-time asymptotic form of $\psi(t)$ underlying the analytic expression for the diffusion coefficient (note, however, that we found it necessary to include some results which supplement the original calculation by Havlin \emph{et al.}). We shall begin by computing the  probability $T_n(\ell)$ that a random walker starting at site $y=1$ takes \emph{at least} $n$ steps along a tooth of length $\ell$ before arriving for the first time at the bottom of the tooth ($y=0$, intersection with the backbone). For this purpose, the site $y=0$ can be thought of as a perfect trap, implying that $T_n(\ell)$ can be identified with the survival probability of the walker up to the $n$-th time step given that its initial position is $y(0)=1$.

In order to compute $T_n(\ell)$, we shall choose for convenience a boundary condition that is slightly different from the one employed in the simulations (the latter corresponds to the one considered by Havlin et al.~\cite{Havlin87b}). Unless otherwise specified, throughout the present section we shall assume that a walker at the end of a tooth ($y=\ell$) will always step back to the site $y=\ell-1$ immediately below the end site. Note the difference with respect to the boundary condition implemented in the simulations, which specifies that the walker either remains at the end site $y=\ell$ (if it attempts to ``exit'' the tooth) or else it steps back to site $y=\ell-1$, whereby each of these two mutually exclusive events takes place with probability 1/2.

With our choice for the boundary condition, the cumulative probability $T_n(\ell)$ becomes identical with the survival probability of a walker moving on a one-dimensional lattice with $2\ell$ sites, whereby both end sites play the role of perfect traps. For our purposes, the above setting is equivalent to a walker placed on a \emph{ring} with $2\ell$ sites, i.e., $2\ell-1$ non-absorbing sites and a single perfect trap, whereby the walker's initial position is a site contiguous to the trap.

A similar reasoning applies for the boundary condition chosen by Havlin \emph{et al.} and implemented in our simulations; however, the equivalent ring would consist of $2\ell+1$ rather than $2\ell$ sites. While this difference can be disregarded for large enough $\ell$, it becomes increasingly relevant in the limit of short teeth. For $0<\alpha<1$ it turns out that the statistical weight of long teeth is very relevant, and so the difference in the boundary condition is negligible for the computation of the diffusion coefficient. However, we shall see that this difference can no longer be neglected in the $\alpha>1$ case.

Let $p_n(y)$ be the probability that the walker is at position $y$ at step $n$ when it starts at $y=1$. The boundary conditions then are $p_n(0)=p_n(2\ell)=0$, and the initial condition is $p_0(y)=\delta_{1,y}$. These probabilities satisfy the difference equation
\begin{equation}
p_{n+1}(y)=\frac{1}{2}\, \left[p_n(y-1)+p_n(y+1)\right].
\end{equation}
The corresponding solution is
 \begin{align}
\label{pnyB}
p_n(y)&=\frac{1}{\ell} \sum_{j=1}^{2\ell-1} \cos^n\beta_j \, \sin\beta_j \, \sin(\beta_j y),
\end{align}
where  $\beta_j= \pi j/2\ell$ [It should be noted that the solution reported in Eq.~(A4) in Ref.~\cite{Havlin87b} is inconsistent with the initial condition $p_0(y)=\delta_{1,y}$]. The survival probability is then given by the expression $T_n(\ell)= \sum_{y=0}^{2\ell} p_n(y)$. For large $\ell$ one can perform the approximations
$\cos^n \beta_j \sim \exp[ -n \beta_j^2/2 ]$ and $\sin\beta_j \, \sin(\beta_j y) \sim \sin^2(\pi j/2)$, which lead to the following asymptotic approximation for $T_n(\ell)$:
\begin{equation}
\label{TnBasyn}
T_n^{\text{asy}}(\ell)=\frac{2}{\ell} \sum_{j=0}^{\infty}\exp[  -n \pi^2 (2j+1)^2
/(8\ell^2)].
\end{equation}
The above asymptotic expression does surprisingly well even if $\ell$ is not too large. The agreement with the exact formula for $T_n(\ell)$ is especially good for large $n$. 

The next step in our route to an analytic expression for $\psi(t)$ is the computation of the probability $U_n(\ell)$ for trapping to take place \emph{exactly} at the $n$-th time step. We note that the absorption probability $U_n(\ell)$ can be regarded as the probability to reach either of the perfect traps located at $y=0$ and $y=2\ell$ after exactly $n$ steps (first-passage probability). Clearly, $U_n(\ell)$ can be expressed as a difference between two survival probabilities, namely,
\begin{equation}
\label{UnfromTn}
U_n(\ell)=T_{n-1}(\ell)-T_{n}(\ell),
\end{equation}
where the initial conditions $T_0(\ell)=1$ and $U_0(\ell)=0$ hold. For even values of $n$ the probability $U_n(\ell)$ must vanish, since for a walker starting at $y=1$ it is impossible to reach either trap after an even number of steps. Therefore, for any positive integer $m$ one has $T_{2m-1}=T_{2m}$, implying that the equality $U_n(\ell) = T_{n-2}-T_n$  holds for odd-valued $n$. As a result of this, the large-$n$ asymptotic expression of $U_n$ can be estimated by regarding the time step $n$ as a continuous variable and computing the corresponding derivative, i.e.,
\begin{align}
\label{UnDerTn}
U_n^{\text{asy}}(\ell)= -2  \;  \frac{d T_{n}^{\text{asy}}(\ell)}{dn},\,\, n \,\, \text{odd}.
\end{align}
In order to make further progress, we must now find estimates of the survival probability $T_n(\ell)$ and the absorption probability $U_n(\ell)$ averaged over an ensemble of teeth with different lengths. To this end, we shall use the approximations $T_n^{\text{asy}}(\ell)$ and $U_n^{\text{asy}}(\ell)$ as defined above. We shall distinguish two cases, namely, $0<\alpha<1$ and $\alpha>1$. In the former case, the mean first-passage time to the intersection with the backbone does not exist, whereas in the latter case it is a finite quantity which will later prove useful for the computation of the diffusion coefficient.

\subsection{Case $0<\alpha<1$}
\label{subs:alphapos}
The average value of $T_n(\ell)$ with respect to the tooth length distribution is
\begin{equation}
\langle T_n \rangle\equiv \langle T_n(\ell)\rangle = \sum_{\ell=1}^\infty {\mathcal P}(\ell) T_n(\ell),
\end{equation}
where $\mathcal{P}(\ell)$ denotes the probability that a randomly chosen tooth has a length of \emph{exactly} $\ell$ units. For the special case of the long-tailed distribution described by Eq.~\eqref{Pl}, the large-$n$ behavior of the survival probability is well described by the following approximation:
\begin{equation}
\label{avTup}
 \langle T_{n}^{\text{asy}}(\ell)\rangle= \alpha \,\int_0^\infty d\ell \, \ell^{-1-\alpha} \, T_{n}^{\text{asy}}(\ell). 
\end{equation}
Using Eq. \eqref{TnBasyn} gives 
\begin{equation}
\label{avT}
 \langle T_{n}^{\text{asy}}(\ell)\rangle=\frac{2\hat\psi_\alpha}{1+\alpha} \; n^{-(1+\alpha)/2}, 
\end{equation}
where we have set 
\begin{equation}
\label{Talpha}
\hat\psi_\alpha=2^{ (1+\alpha)/2} \left(2^{\alpha +1}-1\right)
   \pi ^{-\alpha -1} \alpha \, \Gamma \left(\frac{\alpha+3}{2}\right) \zeta (\alpha
   +1).
\end{equation}
In the above expression, $\zeta(\cdot)$ stands for the Riemann zeta function. Note that the lower limit in the integral appearing in  Eq.~\eqref{avTup} has been shifted from 1 to 0 in order to simplify its analytical evaluation. This approximation is safe, since for $\ell$-values between 0 and 1 and large $n$ the integrand becomes vanishingly small  due to the fast decay of $T_{n}^{\text{asy}}(\ell)$ for sufficiently small values of $\ell$ (cf. Eq. ~\eqref{TnBasyn}). Using the asymptotic form \eqref{avT} in Eq.~\eqref{UnDerTn} one finds
\begin{equation}
\label{EqUnAsy}
 \langle U_{n}^{\text{asy}}(\ell)\rangle= 2 \hat\psi_\alpha \; n^{-(3+\alpha)/2}
\end{equation}
for odd-valued  $n$ satisfying $n\gg 1$.

Let $\psi_n(\ell)$ be the probability that a walker having started its walk at the backbone site $(x,y=0)$ returns to it after exactly $n-1$ time steps, and then hops to a contiguous backbone site $(x\pm 1,y=0)$ at the $n$-th time step. Both sites $(x\pm 1,y=0)$ can be regarded as perfect traps, implying that the quantity $\psi_n(\ell)$ can be interpreted as a first-passage probability to either of the two traps.

\begin{figure} [thb]
  \begin{center}
\vspace*{-0.1cm}
\includegraphics[width=0.5\textwidth, angle=0]
{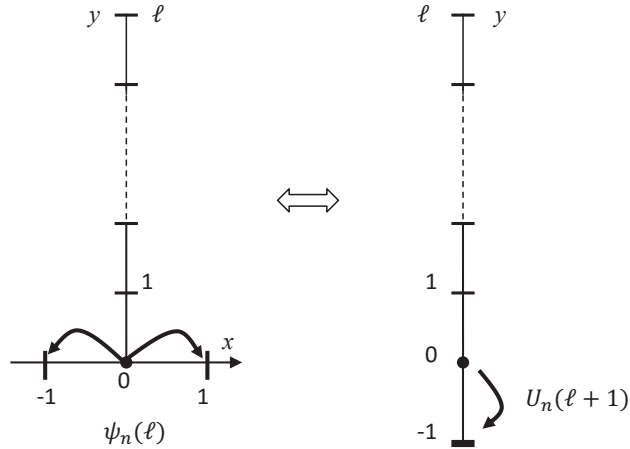}
\vspace*{-0.2cm}
\caption{
\footnotesize
\baselineskip 0.3cm
Graphical representation of the equivalence between $\psi_n(\ell)$ and $U_n(\ell+1)$. The
probability $1/2$  of the walker jumping from $(x=0,y=0)$ to \emph{either} $(x=+1,y=0)$ or $(x=-1,y=0)$ in the left figure is equal to the probability of the walker jumping from $y=0$ to $y=-1$ along the extended tooth shown on the right figure.}
\label{FigPsiU}
  \end{center}
\end{figure}

Let us recall that the probability $\theta$ for the walker to perform a transition $(x,y=0)\rightarrow (x\pm 1,y=0)$ along the backbone has been assumed to be the same ($\theta=1/2$) as the probability to move upwards $(x,y=0)\rightarrow (x,y=1)$. As a result of this prescription, one can see, upon a bit of reflection,  that the  probability $\psi_n(\ell)$ of absorption at sites $(x\pm 1,y=0)$ given that the walker starts at site $(x,y=0)$ is equal to the probability $U_n(\ell+1)$ of absorption at $y=-1$ given that the walker starts at site $y=0$ on a one dimensional lattice stretching from $y=-1$ to $y=\ell$ (see Fig.~\ref{FigPsiU}). Hence, 
\begin{equation}
\label{Upsi}
\psi_n(\ell)= U_n(\ell+1).
\end{equation}

\begin{figure} [thb]
  \begin{center}
\vspace*{-0.1cm}
\includegraphics[width=0.5\textwidth, angle=0]
{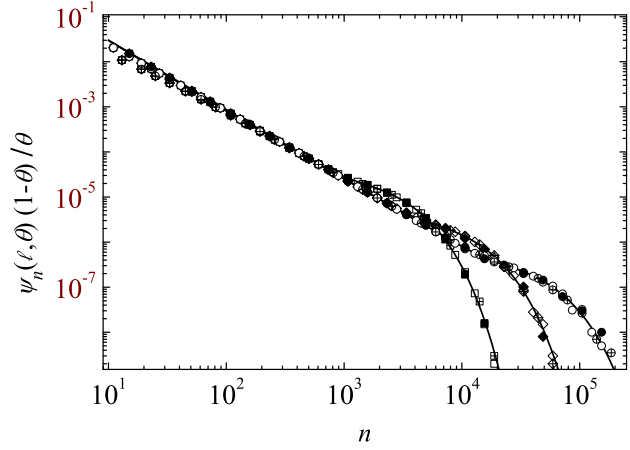}
\vspace*{-0.2cm}
\caption{
\footnotesize
\baselineskip 0.3cm
Log-log plot of simulation results for $\psi_n(\ell,\theta)\,(1-\theta)/\theta$ vs. $n$ for $\ell=50$ (squares), $\ell=100$ (diamonds) and $\ell=200$ (circles) for $\theta=1/2$ (open symbols), $\theta=1/3$ (solid symbols) and $\theta=2/3$ (crossed symbols) for $10^9$ realizations.   Note the excellent collapse of the simulation results for the three different values of $\theta$ and sufficiently large $n$ ($n \ge 50$, say).  The solid curves are the asymptotic values of $\psi_n(\ell)$  obtained by means of Eqs.~\eqref{TnBasyn}, \eqref{UnDerTn} and \eqref{Upsi}. }
\label{Figpsin}
  \end{center}
\end{figure}

When $\theta\neq 1/2$, the corresponding probability $\psi_n(\ell,\theta)$ differs from $\psi_n(\ell)\equiv\psi_n(\ell,\theta=1/2)$. However, in the large-$n$ regime there is a simple relation between both quantities, namely,  
\begin{equation}
\label{psintheta}
\psi_n(\ell,\theta)=\frac{\theta}{1-\theta}\,\psi_n(\ell).
\end{equation}
The proof of the above equation is given in Appendix A.  In addition, Eq. ~\eqref{psintheta}
is confirmed by the numerical results displayed in Fig.~\ref{Figpsin}.

From the definition of $\langle T_{n}^{\text{asy}}(\ell)\rangle$ given by Eq.~\eqref{avT}
one can find the relation $\langle T_{n}^{\text{asy}}(\ell+1)\rangle=\langle T_{n}^{\text{asy}}(\ell)\rangle [1+O(1/n^{1/2})]$ straightforwardly. Thus, setting $\psi_n(\ell)\approx U_n^{\text{asy}}(\ell)$, taking the average over the tooth length pdf $P(\ell)$, and using Eq.~\eqref{EqUnAsy} one obtains
\begin{equation}
\label{psin4}
\langle \psi_n\rangle \sim 2  \hat \psi_\alpha\; n^{-(3+\alpha)/2},\quad n\gg 1, \;\; n  \quad \text{odd}.
\end{equation}

For $\theta\neq 1/2$, this formula would be the same, except for the fact that $\hat \psi_\alpha$ should be replaced with  $\theta \hat \psi_\alpha/(1-\theta)$ in that case. 

We now seek to establish a relation between the above discrete-time description in terms of the probabilities $\psi_n$ and the continuous-time description based on the corresponding waiting time pdf $\psi(t)$. To this end, we first average $\psi_n$ over odd- and even-valued times $n$, i.e., we take
\beq
\label{psit}
\psi(t)\Delta t\approx \psi(t)\times 2= \langle \psi_{n-1}\rangle + \langle \psi_n\rangle.
\eeq
Eq.~\eqref{Upsi} together with the fact that $U_n$ vanishes for even values of $n$ implies that $\langle \psi_n\rangle=0$ for even-valued $n$.  Taking this into account and making use of Eq.~\eqref{psit} in \eqref{psin4} we find
\begin{equation}
\label{psioft}
\psi(t)\sim \hat \psi_\alpha\; t^{-(3+\alpha)/2},\quad t\gg 1.
\end{equation}
We are now in the position to perform an explicit computation of the diffusion coefficient.
To this end, we recall that a CTRWer whose motion is described
by the asymptotic long-tailed waiting time pdf
\begin{equation}
\label{psig}
 \psi(t)\sim\frac{\gamma \tau^\gamma}{\Gamma(1-\gamma)}\; t^{-1-\gamma},  \quad
t \quad \text{large},
\end{equation}
with $0<\gamma<1$, displays subdiffusive behavior provided that its step length distribution has a finite variance $\Sigma^2$, implying that its MSD can be written as follows:
\begin{equation}
\label{MSDt}
\langle x^2\rangle \sim \frac{2K_\gamma}{\Gamma(1+\gamma)}\; t^\gamma=D_0(\gamma) t^\gamma,
\end{equation}
where $K_\gamma \equiv \Sigma^2/(2\tau^\gamma)$ and $\tau$ is a characteristic time scale for the waiting time between jumps \cite{Metzler00}. Comparing  Eq.~\eqref{psioft} with Eq.~\eqref{psig} one finds
$\gamma=(1+\alpha)/2$ and $\tau^\gamma  = \Gamma(1-\gamma)\; \hat \psi_\alpha/{\gamma}$. Inserting these expressions into Eq.~\eqref{MSDt}
we find
\begin{equation}
D_0(\alpha)=\frac{1+\alpha}{2\, \Gamma(\frac{3+\alpha}{2})\Gamma(\frac{1-\alpha}{2}) \hat \psi_\alpha}\, \Sigma^2.
\end{equation}
The above equation can be further simplified by Eq.~\eqref{Talpha}:
\begin{equation}
\label{D0alpha4}
D_0(\alpha)=\frac{2^{-(1+\alpha)/2} \,\pi ^{\alpha} \,\cos \left(\frac{\pi  \alpha }{2}\right)}{\alpha  \left(2^{\alpha +1}-1\right)  \Gamma \left(\frac{\alpha +3}{2}\right) \zeta (\alpha +1)} \, \Sigma^2.
\end{equation}
If $\theta\neq 1/2$, the diffusion coefficient is simply $D_0(\alpha,\theta)=(1-\theta) D_0(\alpha)/\theta$. In our case, Eq.~\eqref{D0alpha4} is further simplified by taking into account that the motion along the backbone proceeds by transitions between nearest-neighbor sites (occasionally delayed by excursions along the teeth). Thus, one must take $\Sigma^2=1$.  The analytic expression provided by Eq.~\eqref{D0alpha4} turns out to be in excellent agreement with our simulation data (cf. Fig.~\ref{D0}).  We note that changes in the value of  $\Sigma$ can be interpreted as a change in the density of teeth along the backbone, which according to Eq.~\eqref{D0alpha4} has an influence on the diffusion coefficient, but has no effect on the diffusion exponent $\gamma$. This differs from the results reported for the dendritic system studied in Ref.~\cite{Santamaria2011}.

\subsection{Case $\alpha>1$}
\label{alphaGT1}
Here, it is well known that normal diffusion takes place regardless of the value of $\alpha$, i.e., $\gamma=1$ \cite{Havlin87,Havlin87b}. The diffusion coefficient for this case has already been calculated with a variety of different methods \cite{Balakrishnan1995,Revathi1996}. For the sake of completeness, a simple alternative derivation is given below. Our derivation exploits the fact that for $\alpha>1$ the spatial average of the mean dwelling time of the random walker inside a tooth is \emph{finite}, implying that the diffusion coefficient can be written as follows:
\begin{equation}
\label{DifCoe4}
K_1(\alpha)=\frac{\Sigma^2}{2\tau}   \quad \Rightarrow \quad
D_0(\alpha)=\frac{\Sigma^2}{\tau},
\end{equation}
where $\tau$ is the (spatially averaged) mean waiting time between consecutive jumps along the backbone and, in the present context, $\Sigma^2$ is the variance of the distance between consecutive teeth.  Let us now introduce the quantity  $\tau(\ell)$ as the average number of time steps that it takes for a walker initially located at the bottom of a tooth of $\ell$ units to jump along the backbone, that is, to perform the transition $(x,y=0)\to (x\pm 1,y=0)$. One then has
\begin{equation}
\label{tauasave}
\tau=\langle\tau(\ell)\rangle_\ell=\sum_{\ell=1}^\infty   \mathcal{P}(\ell)\, \tau(\ell).
\end{equation}
Next, let us denote by $t_R(\ell)$ the average number of time steps required by a walker initially located at $(x,y=0)$ to return to its initial position given that it starts moving vertically along the tooth. A walker starting at $(x,y=0)$ can reach $(x\pm 1,y=0)$ after one time step with probability $1-\theta=1/2$ provided that it does not enter the tooth, or it may enter the tooth once with probability $\theta=1/2$, come back to $(x,y=0)$ after $t_R(\ell)$ time steps, and then perform the final transition $(x,y=0)\to (x\pm 1,y=0)$ with probability $1-\theta=1/2$, and so on. Summing up the contributions from trajectories involving a different number of returns to the intersection with the backbone, one finds
\begin{align}
\label{tauL}
\tau(\ell)&=1\,\times (1-\theta)+(t_R(\ell)+1)\,\theta (1-\theta)+(2 t_R(\ell)+1)\,\theta^2(1-\theta)
+\ldots \nonumber \\
&=  (1-\theta)\sum_{n=0}^\infty \theta^n (n\, t_R(\ell)+1)=
1+\frac{\theta}{1-\theta}\,t_R(\ell).
\end{align}
In writing the above equation, we have taken into account that the time needed for the walker to move by one lattice spacing [i.e., from $(x,y=0)$ to either of its two nearest neighbor sites $(x=x\pm 1, y=0)$] had been chosen to be equal to one.  Had this time been set equal to a different value $t_b$, the expression $nt_R(\ell)+1$ in Eq.~\eqref{tauL} should have been replaced with  $nt_R(\ell)+t_b$.

As already mentioned, for the boundary condition used in the simulations (cf. Secs. II and III), a tooth of length $\ell$ is equivalent to a one dimensional periodic lattice with $2\ell+1$ sites. In that case, $t_R(\ell)$ can be understood as the mean return time to the origin of a  walker on a ring of length $2\ell+1$. On the other hand, it is well known that for a periodic $N$-site lattice this return time is precisely identical with the number of lattice sites $N$ \cite[Eq.~(4.172a)]{Weiss1994}. In the above setting, we have $N=2\ell+1$, leading to $t_R(\ell)=2\ell+1$.  From Eqs. \eqref{tauL} and \eqref{tauasave} we then find
\begin{equation}
\tau=\frac{\theta}{1-\theta}+\frac{2\theta}{1-\theta}\sum_{\ell=1}^\infty \mathcal{P}(\ell)\, \ell\equiv \frac{1+2\theta \langle \ell\rangle}{1-\theta},
\end{equation}
in agreement with the results obtained in Refs.~\cite{Balakrishnan1995} and \cite{Revathi1996} [see, e.g., Eq.~(48) in ref.~\cite{Balakrishnan1995}].  For the particular case where ${\cal P}(\ell)$ is given by Eq.~\eqref{PLdiscreta}, one finds $\langle \ell \rangle=\zeta(\alpha)$, leading to the equation $\tau=2+2\zeta(\alpha)$ for $\theta=1/2$. As a result of this, one has 
\begin{equation}
\label{D0alphagt1}
D_0(\alpha)=\frac{\Sigma^2}{2+2\zeta(\alpha)}.
\end{equation}
As in the $\alpha<1$ case, the agreement of the above results with the simulation results displayed in Fig.~\ref{D0} is excellent. Note that our choice to restrict the displacements along the backbone to nearest-neighbor jumps between sites separated by one lattice spacing implies that one must take $\Sigma^2=1$.  In the case of arbitrary $\theta$, Eq.~\eqref{D0alphagt1} must be replaced with the more general expression 
\beq
D_0(\alpha)=\frac{1-\theta}{1+2\theta \zeta(\alpha)}\,\Sigma^2.
\eeq
Finally, we can easily extend our results to the case where the boundary condition is the one used in Sec.~IV [recall that in this case a walker at the end of a tooth ($y=\ell$) will always step back to site $y=\ell-1$]. As already mentioned there, this boundary condition corresponds to treating a tooth of length $\ell$ like a ring with $2\ell$ sites. Correspondingly, one has $t_R(\ell)=2\ell$ and $\tau=1+2\langle \ell \rangle$, leading to
$D_0(\alpha)= \Sigma^2/[1+2\zeta(\alpha)]$, or to $D_0(\alpha)= \Sigma^2/[1+2\theta (1-\theta)^{-1}\zeta(\alpha)]$ for $\theta\neq 1/2$.

\subsection{Case $\alpha=1$}
For $\alpha=1$, the waiting time pdf given by \eqref{psioft} is simply $\psi(t)\sim \hat \psi_1\; t^{-2}$, where $\hat \psi_1=1$ [or $\hat \psi_1=\theta/(1-\theta)$]. For this specific form of waiting time pdf it is known that the MSD
behaves as follows   \cite{Shlesinger1974}:
\beq
\langle x^2(t) \rangle \sim \frac{\Sigma^2}{\hat \psi_1} \frac{t}{\ln(t)}.
\eeq
This result is again confirmed by our numerical simulations (see the bottom curve in Fig.~\ref{Xt}).

\section{\fat Influence of binding/unbinding kinetics }
Retardation effects associated with binding/unbinding kinetics in biological or biomimetic systems have been widely studied, notably by means of Monte Carlo simulations (see e.g. Refs.~\cite{Saxton96} and \cite{Saxton07}). In what follows, we shall study such effects for the particular case of the random comb model.

In our simulations, we implemented binding/unbinding processes as follows. At a given time, walkers at any site of the comb could be found in either of two states, namely, ``bound'' or ``unbound''. Our collection of walkers was initially distributed at random along the $x$-axis, and all of them were initially unbound.  Subsequent transitions between the unbound state and the bound state proceeded as follows. Whenever an unbound walker would jump to a nearest neighbor site, it would bind to it with probability $\kon$ (the time step was taken to be unity, thereby allowing one to interpret $\kon$ as a rate constant).  In turn, walkers in the bound state could unbind with probability $\koff$ (rate constant for unbinding processes) when selected by the simulation algorithm, and they were subsequently free to jump to a nearest neighbor site on the discretized comb.

At this stage, a comment on the physical origin of the above rate constants is in order. At a  mesoscopic level, one may regard $\kon$ and $\koff$ as effective parameters whose values are in principle obtainable from experiments. However, one should bear in mind that a more microscopic picture would bring molecular potentials between the diffusing molecules (``the walkers'') and the biological matrix (``the comb'') into play. The latter approach is beyond the scope of the present work and will not be further discussed here.

\begin{figure} [thb]
  \begin{center}
\vspace*{-0.1cm}
\includegraphics[width=0.5\textwidth, angle=0]
{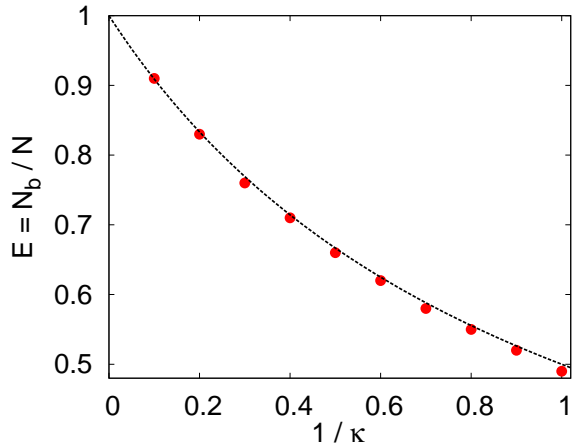}
\vspace*{-0.2cm}
\caption{
\footnotesize
\baselineskip 0.3cm
Average fraction of bound particles $E \equiv N_b / N$ as a function of $\kappa$.
}
\label{KK-E}
  \end{center}
\end{figure}

The fraction of bound particles $E=N_b/N$ can be interpreted as the normalized ``binding energy'' $E$ of the system.  From the analysis of the corresponding kinetic equations this fraction is expected to be $E \sim \kon / (\kon + \koff)$ in the long-time limit. As a result of this, the energy should follow the law $E = \kappa/(1+\kappa)$, where $\kappa \equiv \kon / \koff$ is the ratio of rate constants. This is in full agreement with the simulation data displayed in Fig.~\ref{KK-E} (see dots). Note that the value of $E$ is independent of the exponent $\alpha$.

\begin{figure} [thb]
  \begin{center}
\vspace*{-0.1cm}
\includegraphics[width=0.5\textwidth, angle=0]
{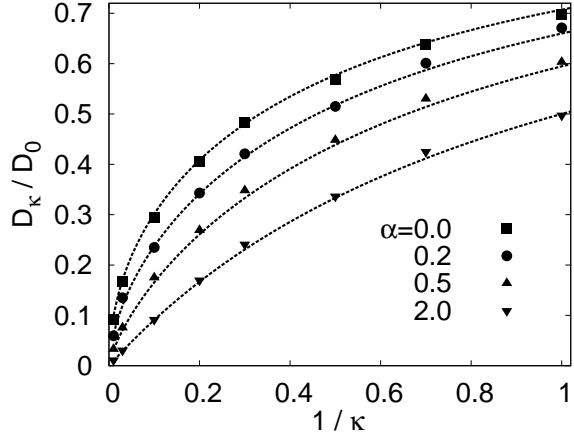}
\vspace*{-0.2cm}
\caption{\footnotesize
\baselineskip 0.3cm
Normalized diffusion coefficient $D_{\kappa}/D_0$ versus $\kappa$ [$D_0$ is the diffusion coefficient for $\kon=0$]. The dashed curves represent the theoretical prediction given by Eq.~\eqref{eq:DK}, whereas the symbols represent simulation results. 
\label{KK-DK}
}
  \end{center}
\end{figure}

We now proceed to quantify the impact of the delay introduced by binding/unbinding events on the diffusion coefficient for the walker motion along the backbone. The MSD in the long time regime $\langle x^2(t) \rangle = D_{\kappa}(\alpha) t^{\gamma}$ defines the $\kappa$-dependent diffusion coefficient. Numerical results for the normalized effective diffusion coefficient $D_{\kappa} / D_0$ are shown in Fig.~\ref{KK-DK}. Here, the effective diffusion coefficient $D_0=D_0(\alpha)$ is the one defined previously for the case where binding-unbinding events are absent. As expected, the diffusion coefficient is seen to decrease with increasing $\kappa$, i.e., with increasing $\kon$. This can be explained as follows. A walker arriving at a new location either remains in the unbound state with probability $1-\kon$ (and is thus free to jump to a nearest neighbor site), or else it becomes bound with probability $\kon$. In the first case, the time required by the walker to move to a neighboring site is one (in our units), whereas in the latter case the average waiting time due to binding is $\koff^{-1}+1$ (i.e., $\koff^{-1}$ time units to escape the binding state and one additional time unit to perform the nearest neighbor jump). The average time between consecutive nearest neighbor jumps then becomes $(1-\kon)\cdot 1+\kon\cdot (\koff^{-1}+1)=1+\kappa$. Therefore, for the case with binding, the time required to perform a transition between neighboring sites is increased by the factor $(1+\kappa)$  with respect to the case without binding. Hence, the MSD $\langle x^2(t) \rangle = D_{0} t^{\gamma}$ along the backbone becomes  $\langle x^2(t) \rangle = D_{0} [t/(1+\kappa)]^{\gamma}$ in the presence of binding, and consequently,
\beq
\frac{D_{\kappa}}{D_0} = \frac{1}{(1+\kappa)^{\gamma}} = \left(\frac{E}{\kappa}\right)^{\gamma}.
\label{eq:DK}
\eeq
In Fig.~\ref{KK-DK}, the plots of the analytical expression given by Eq.~\eqref{eq:DK} (dashed curves) can be seen to be in excellent agreement with numerical data from MC simulations results.

It should be noted that Eq.~\eqref{eq:DK} can be understood as the extension to the anomalous diffusion case of the computation for the effective diffusion coefficient in the presence of binding/unbinding processes that can be found in the literature for the standard diffusion case; see, e.g., Eq.~\eqref{eq:DK} in Ref.~\cite{McNally04}, which is recovered from Eq.~ \eqref{eq:DK}
when $\gamma=1$. The effective diffusion regime mentioned in Ref.~ \cite{McNally04} turns out to be dominant in comb-like systems with $\alpha<1$, since it corresponds to the case where the binding/unbinding reactions occur on a much shorter time scale than transport along the backbone.

Finally, we also note that from the formula \eqref{eq:DK} one finds $\kappa=(D_0/D_\kappa)^{1/\gamma}-1$, implying that in the subdiffusive regime $0<\alpha<1$ knowledge of the normalized diffusion coefficient is not enough to infer the ratio of rate constants characterizing the binding/unbinding processes (to calculate $\kappa$, one must additionally know the value of $\alpha$). This is a key difference with respect to the normal diffusion regime with $\alpha\ge 1$.

Conversely, in experiments where the parameters for the binding/unbinding processes do not change, the value of the exponent $\alpha$ characterizing the comb geometry can only be determined if $\gamma \le 1$. In such cases, knowledge of  $D_\kappa/D_0$ would still be insufficient, since the value of $\kappa$ is also needed. 

\section{FRAP dynamics on the random comb}
\label{secFRAP}
Both the diffusion coefficient $D_0(\alpha)$ calculated in Sec.~IV and its corrected value in the presence of binding/unbinding events are expected to be useful for quantitative studies aiming at the characterization of diffusive transport in comb-like biological structures, and notably in spiny dendrites \cite{Schmidt2003, Schmidt2007, Santamaria06, Santamaria2011, Byrne2011}. The ultimate goal is the comparison with experiments where representative quantities depending on transport properties are monitored.

In the above context, FRAP experiments are a widely used technique to explore binding interactions of membrane proteins in cells. For a comprehensive, up-to-date review on FRAP and other microscopy techniques the reader is referred to \cite{Hoefling13}. As already mentioned in the introduction, in FRAP experiments particles are stained with a fluorescent dye and then those in a small region (the ``bleached spot'') are photobleached with a laser pulse. Following this,  one measures the fluorescent signal recovery as the bleached spot is progressively filled with particles diffusing from the region outside the spot.

FRAP techniques are especially well-suited when transport processes are very slow and, when a significant portion of molecules is immobile, they appear to be more robust than other fluorescence-based techniques \cite{Hoefling13}. Nowadays, FRAP experiments are widely used to characterize \emph{in vivo} protein motion \cite{Reits2001}, diffusion-controlled drug delivery \cite{Meyvis} and morphogen transport \cite{Muller2013}. In these systems, diffusing proteins in the cell nucleus bind reversibly to the immobile nuclear structure. The cell nucleus may be considered to be a `crowded' environment causing the proteins to move subdiffusively on sufficiently long time scales.

From a theoretical point of view, a method aimed at reproducing FRAP recovery curves by means of coupled reaction-diffusion equations was first developed in Ref.~\cite{McNally04}
and subsequently generalized in follow-up works \cite{McNally05, McNally10}. In Ref.~\cite{Yuste2014}, the solutions corresponding to a two-dimensional geometry and a circular bleached spot were extended to the case where the diffusing particles perform a subdiffusive CTRW rather than standard Markovian walks. Interestingly, experimental recovery curves previously described by a normal diffusion model were found to be equally well fitted by a fractional diffusion equation arising from the CTRW model.

In the present section, we shall study the FRAP phenomenology in the random comb model. To this end, we initially placed a collection of particles on the backbone, eliminated those of them within a segment of the backbone (the bleached spot) and subsequently let the remaining ones perform random walks according to the simulation procedure described below. The time evolution of the number of unbleached particles inside the bleached spot and the associated concentration recovery curves were computed numerically and shown to be reproducible by means of fractional diffusion equations underlying the corresponding one-dimensional CTRW model. The mathematical treatment of the latter is similar to the one used in Ref.~\cite{Yuste2014} for the two-dimensional case.

\subsection{Simulation procedure}
The simulations for the numerical computation of FRAP recovery curves were performed as follows. We first defined a discretized comb with randomly distributed tooth length in $y$-direction and a backbone in $x$-direction consisting of $M$ lattice sites subject to periodic boundary conditions, whereby the lattice spacing in $x$- and $y$-directions was chosen to be the same. Following this, $N$ non-interacting walkers were randomly scattered along the backbone, and all the walkers found within a segment of length $L<M$ lattice sites were then ``bleached'', i.e., removed from the system.

Following this, we let the walkers perform nearest-neighbor jumps and thereby spread throughout the entire comb structure. We then monitored the time evolution of the average number of walkers $N_{spot}(t)$ found within a spot of size $L$ or, more conveniently, the normalized average number of walkers $C_L(t)$ dwelling inside the spot at time $t$:

\begin{equation}
C_L(t) = \frac{\langle N_{spot}(t)\rangle }{\langle N_{spot}(\infty)\rangle}.
\label{eq:Ct}
\end{equation}
The normalizing quantity $\langle N_{spot}(\infty)\rangle$ was easily obtained by taking into account that the mean number of particles per unit length in the final homogeneous state is the same as immediately after the photobleaching, that is, $(N - \langle N_{bl} \rangle)/M = \langle N_{spot}(\infty)\rangle/L$, where $N_{bl}$ is the number of bleached particles. Thus, one finds $\langle N_{spot}(\infty)\rangle= (N-\langle N_{bl}\rangle)(L/M)$.

\subsection{Analytical and numerical results}
The time evolution of $C_L(t)$ can be studied both analytically and numerically. Our subsequent analysis relies on the CTRW approach used in Sec.~ IV to successfully analyze diffusion on the random comb. We have shown that particle spread along the backbone can be effectively described by one-dimensional diffusion of CTRWers with a waiting time density $\psi(t)$ given by Eq.~\eqref{psioft} or, equivalently, by Eq.~\eqref{psig}. On the other hand, it is well known that in the long-time limit the evolution of the concentration $c(x,t)$ of such CTRWers obeys the following fractional diffusion equation \cite{Metzler00}:
\begin{align}
 \frac{\partial}{\partial t} c(x,t) &= K_\gamma~_{0}D_t^{1-\gamma} \frac{\partial^2}{\partial x^2} \,c(x,t),
 \label{rlfde}
\end{align}
where ${}_{0}D_t^{1-\gamma}$ stands for the so-called Riemann-Liouville fractional derivative.

Let us denote by $c_0$  the value  of the concentration  $c(x,t)$ before the bleaching. In what follows we assume that our comb system extends from $x=-M/2$ to $x=M/2$, whereby the bleached spot extends from $x=-L/2$ to $x=L/2$. We consider the case of perfect bleaching described by the initial condition \cite{McNally04}
\beq
c(x,0)=\begin{cases}
0, & |x| \le L/2,\\
c_{0}, &  L/2 < |x| \le M/2.
\end{cases}
\label{initcond}
\eeq
In the simulations, periodic boundary conditions are taken, i. e., $c(x,t)=c(x+L,t)$.  The solution to the boundary value problem posed by Eqs. \eqref{rlfde}- \eqref{initcond} and the periodic boundary condition is easily found by means of the Green's function method  (see Appendix B). One finally obtains
\beq
\widetilde{c}(x,s)=\begin{cases}
\displaystyle \frac{c_{0}}{s}\frac{e^{q_\gamma [M-(L/2)]}-e^{q_\gamma L/2}}{e^{q_\gamma M}-1} \,
\cosh(q_\gamma x),  & |x| \le L/2, \\[0.2cm]
\displaystyle
\frac{c_{0}}{s}\left[1-\sinh\left(q_\gamma \frac{L}{2}\right) \frac{e^{q_\gamma x}+e^{q_\gamma (M-x)}}{e^{q_\gamma M}-1}  \right], & |x|>L/2,
\label{f-conc}
\end{cases}
\eeq
where $q_\gamma =\sqrt{s^\gamma/K_\gamma}$.

The Laplace transform of the spatial average of $c$ over the bleached spot is
\beq
\langle \widetilde{c} \rangle = \frac{2}{L} \int_{0}^{\frac{L}{2}} \widetilde{c}(x,s) \, dx.
\eeq
This gives
\beq
\langle \widetilde{c} \rangle (s)= \frac{c_{0}}{s}\frac{1+e^{q_\gamma M}-e^{q_\gamma L}-
e^{q_\gamma (M-L)}}{(e^{q_\gamma M}-1) q_\gamma L}.
\label{finitesizeFRAP}
\eeq

Let $c^\star=c(x,\infty)$ be the final particle concentration. Since $\langle N_\text{spot}(t)\rangle =\langle c\rangle L$ and $\langle N_\text{spot}(\infty)\rangle = c^\star L$, the normalized number of particles inside the bleached spot $C_L(t)$ as defined by Eq.~\eqref{eq:Ct} can be rewritten as $C_L(t)=\langle c\rangle /c^\star$. It is clear that  $ c^\star=c_0\, (M-L)/M$, and hence
\beq
\widetilde{C}_L(s)= \frac{M}{M-L}\frac{1+e^{q_\gamma M}-e^{q_\gamma L}-e^{q_\gamma (M-L)}}{(e^{q_\gamma M}-1)
s \, q_\gamma L}.
\label{LaplaceFRAP}
\eeq

\subsubsection{Infinite system}

In the limit $M\to \infty$, Eq.~\eqref{LaplaceFRAP} becomes
\beq
\widetilde{C}_L(s)= \frac{1-e^{-q_\gamma L}}{ s q_\gamma L},
\label{LapCMInf}
\eeq
which can be inverted analytically  to obtain
\beq
C_L(t) = \frac{\left(K_\gamma t^\gamma\right)^{1/2}}{\Gamma(\gamma/2 + 1) L}\,-\,H_{11}^{10}\left[\frac{L}{\left(K_\gamma t^\gamma\right)^{1/2}} \left|
\begin{array}{ll}  (1, \gamma/2)\\
                  (-1,1)  		
                   \end{array}
                   \right.
                    \right]
\label{eq:Cttheory}
\eeq
for $\gamma<1$ (i.e., for $0<\alpha < 1$) and
\beq
C_L(t) = \textrm{erfc}\left(\frac{L}{2\,\left(K_\gamma t^\gamma\right)^{1/2}}\right) + \frac{2\,\left(K_\gamma t^\gamma\right)^{1/2}}{\pi^{1/2} L}\left(1 - e^{-L^2/(4 K_\gamma t^\gamma)}\right)
\label{eq:Cttheory2}
\eeq
for $\gamma=1$ (i.e., for $\alpha \ge 1$). In Eq.~\eqref{eq:Cttheory}, $H_{11}^{10}$ stands for a particular class of Fox's H-function \cite{MathaiBook, Metzler00}. In passing, we note that the solution for the anomalous diffusion case given by Eq.~\eqref{eq:Cttheory} is related to the solution \eqref{eq:Cttheory2} for the normal diffusive case by means of the subordination principle for the underlying CTRW process \cite{Lubelski2008}.

Let us now examine the early-time behavior of $C_L(t)$. To this end, one can either look up the relevant series expansion of the Fox function or take the limit $s \to \infty$ ($q_\gamma \to \infty$) in Eq.~\eqref{LapCMInf}. We choose the second option. To leading order one finds
\beq
\widetilde{C}_L(s) \sim \frac{1}{ s q_\gamma L}= \frac{K_\gamma^{1/2}}{L} s^{-\gamma/2-1},
\eeq
leading to the short-time behavior
\beq
C_L(t) \sim \frac{K_\gamma^{1/2}t^{\gamma/2}}{\Gamma\left(1+\frac{\gamma}{2}\right)\,L}.
\eeq
In the opposite small-$s$ limit  ($s, q_\gamma \to 0$) one has
\beq
\widetilde{C}_L(s)=\frac{1}{s}\left(1-\frac{q_\gamma L}{2}+\frac{q_\gamma^2 L^2}{6}-\cdots\right)
=s^{-1}-\frac{L}{2K_\gamma^{1/2}}\,s^{\frac{\gamma}{2}-1}+ \frac{L^2}{6K_\gamma}\,s^{\gamma-1}-\cdots,
\eeq
leading to the following long-time behavior:
\beq
\label{infsyslate}
C_L(t)=1-\frac{L}{2\Gamma\left(1-\frac{\gamma}{2}\right) K_\gamma^{1/2}t^{\gamma/2}}+\frac{L^2}{6\Gamma\left(1-\gamma\right) K_\gamma t^\gamma}-\cdots.
\eeq
Thus, to leading order one finds
\beq
[1-C_L(t)]^{-1}= 2\Gamma\left(1-\frac{\gamma}{2}\right) L^{-1} K_\gamma^{1/2}t^{\gamma/2}, \quad t\to \infty.
\label{LeadLargeTimes}
\eeq

\subsubsection{Finite system}
\begin{figure} [thb]
  \begin{center}
\vspace*{-0.1cm}
\includegraphics[width=0.5\textwidth, angle=0]
{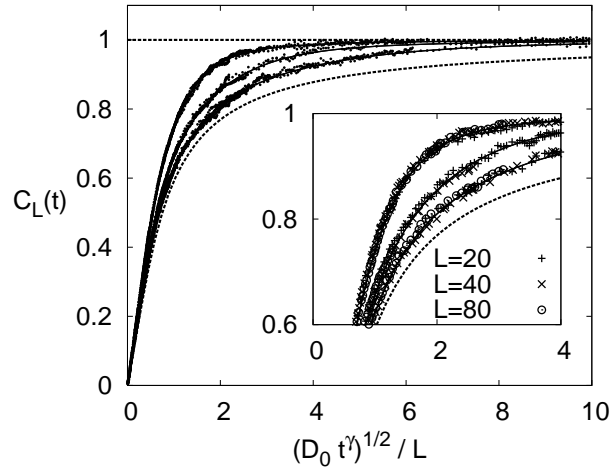}
\vspace*{-0.2cm}
\caption{\footnotesize
\baselineskip 0.3cm
Concentration $C_L(t)$ of particles in the bleached spot as a function of the rescaled time for $\kappa = 0$ for different ratios $L/M$ obtained by fixing the value of $L$ and using three different values of $M$. The $L$-values were set to $20$, $40$ and $80$. Curves for different values of $L$ but for the same value of $L/M$ are seen to collapse into a single curve. The $L/M$-values corresponding to collapsing data sets are, from top to bottom, $0.2, 0.1$ and $0.05$. The dotted recovery curve corresponds to the case of an infinite system [Eq.~\eqref{eq:Cttheory}]. The inset shows a zoom-in of the data in the region where changes in the time derivative of the recovery curve are largest (different symbols are used to distinguish different data sets corresponding to the same value of $L$). The solid curves are plots obtained from the numerical inversion of Eq.~\eqref{LaplaceFRAP}.
}
\label{Ct}
  \end{center}
\end{figure}

\begin{figure} [thb]
  \begin{center}
\vspace*{-0.1cm}
\includegraphics[width=0.5\textwidth, angle=0]
{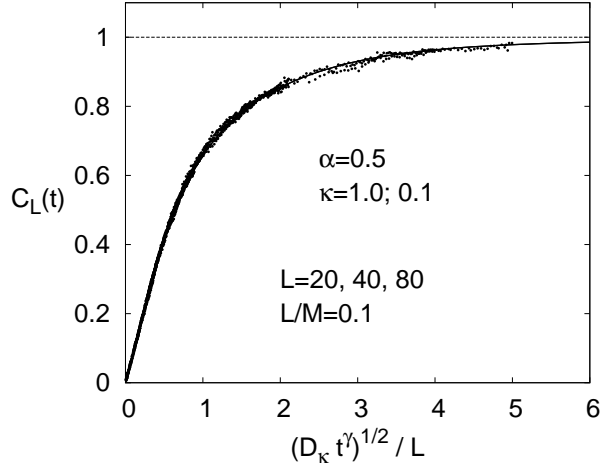}
\vspace*{-0.2cm}
\caption{\footnotesize
\baselineskip 0.3cm
Concentration $C_L(t)$ of particles in the bleached spot as a function of the rescaled time
in the $\kappa > 0$ case. The collapse of data sets corresponding to different values of $\kappa$ is due to the fact that the recovery curves are plotted in terms of the $\kappa$-dependent rescaled time. The solid curve again corresponds to the inversion of Eq.~\eqref{LaplaceFRAP}. }
\label{CtK}
  \end{center}
\end{figure}

Next, we turn to the study of finite size effects. For finite $M$ and $L/M < 1$, the scaling behavior of $C_L(t)$ versus the rescaled time $(D_0 t^{\gamma})^{1/2}/L$ is represented in Figs.~\ref{Ct} and \ref{CtK} for the cases $\kappa=0$ and $\kappa>0$, respectively. As can be seen from Fig.~\ref{Ct}, finite size effects are important in both cases. Both the exponent $\alpha$ and the typical lengths $M$ and $L$ determine the initial slope of the recovery curves. For a given $\alpha$ and a given finite size ratio $L/M$ recovery curves corresponding to different values of  $L$ collapse approximately to a single curve (see Fig.~\ref{Ct}). As shown below, this short-time behavior can also be recovered analytically.

For early times we take the $s\to \infty$ limit of Eq.~\eqref{LaplaceFRAP} and find
\beq
\widetilde{C}_L(s) \sim \left(\frac{M}{M-L}\right) \frac{1}{ s q_\gamma L}= \left(\frac{M}{M-L}\right)\frac{K_\gamma^{1/2}}{L} s^{-\gamma/2-1},
\eeq
i.e.,
\beq
\label{earlytimeEq}
C_L(t) \sim \left(\frac{M}{M-L}\right)\,\frac{K_\gamma^{1/2}t^{\gamma/2}}{\Gamma\left(1+\frac{\gamma}{2}\right)\,L}.
\eeq
In the opposite limit of long times ($s\to 0$), the pertinent expansion yields
\beq
\widetilde{C}_L(s)=s^{-1}-\frac{L(M-L)}{12} K_\gamma^{-1} s^{\gamma-1}+{\cal O}\left(s^{2\gamma-1}\right).
\eeq
In direct space this yields
\beq
\label{latetimeEq}
C_L(t)=1-\frac{L(M-L)}{12 \Gamma(1-\gamma) K_\gamma t^\gamma}+{\cal O}\left(t^{-2\gamma}\right),
\eeq
implying that
\beq
\label{latetimeEq2}
[1-C_L(t)]^{-1}=\frac{12 \Gamma(1-\gamma) K_\gamma t^\gamma}{L(M-L)}+{\cal O}\left(t^{2\gamma}\right).
\eeq
Thus, a log-log plot of the simulation curves for the inverse of the difference between the average normalized concentration and its final value (here normalized to one) as a function of time should yield linear behavior with a slope equal to $\gamma$. Note the difference in behavior with respect to the infinite system [Eq.~\eqref{LeadLargeTimes}], where the growth of $[1-C_{L,\infty}(t)]^{-1}$ is proportional to $t^{\gamma/2}$. This means that the limits $M\to\infty$ and $t\to\infty$ do not commute. 

\subsection{Non-reproducibility of concentration recovery curves with an approach based on scaled Brownian motion}
Diffusion equations with a time-dependent diffusion coefficient (so-called scaled Brownian motion (sBm) in the language of Ref.~\cite{Sokolov2012}) are often used to fit in a rather successful way recovery curves recorded in FRAP experiments \cite{Nagle1992, Lubelski2008, Saxton01}. The idea underlying this ``ad-hoc'' procedure is to replace the diffusion coefficient $D\equiv K_1$ in the standard diffusion equation by a time-dependent expression $D' t^{\gamma-1}$, whereby $D'$ and $\gamma$ are used as fitting parameters. In Ref.~\cite{Thiel2014}, a word of caution is given against the use of such approaches without the corresponding justification at a microscopic level of description. Even though sBm seems to work well in the specific case of FRAP recovery curves, the lack of more detailed information on the elementary transport processes might lead to wrong results if one attempts to compute other quantities.

\begin{figure} [thb]
  \begin{center}
\vspace*{-0.1cm}
\includegraphics[width=0.45\textwidth, angle=0]
{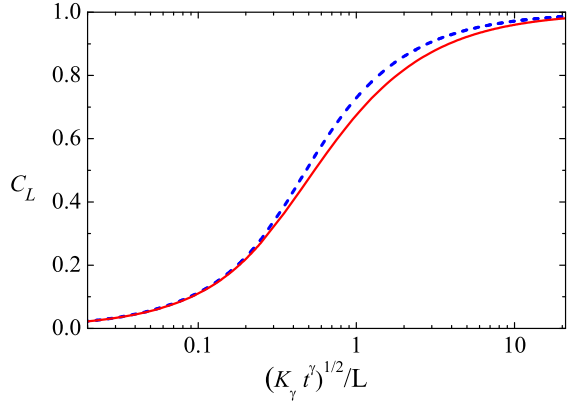}
\caption{\footnotesize
Semi-logarithmic plot representing $C_L(t)$  vs. $\log_{10} (K_\gamma t^\gamma)^{1/2}/L$ for the two extreme cases of a comb model with an infinitely long backbone ($M\to\infty$), namely, (i) the case where all the teeth are infinite, i.e., the case with $\alpha=0$ (implying $\gamma=1/2$, see solid curve),  and (ii) the case where all the teeth are finite, i.e., the case of normal diffusion with $\alpha>1$ (implying $\gamma=1$, see dashed curve) and $K_1\equiv D$.}
\label{CLvsxiL}
  \end{center}
\end{figure}

How well does sBm work for our comb model? Since our CTRW-based exact solution is valid for arbitrary values of $\gamma$, we can answer this question easily. In Fig.~\ref{CLvsxiL} we show a lin-log plot of $C_L$ as a function of $(K_\gamma t^\gamma)^{1/2}/L$ for $\gamma=1/2$ and also for the normal diffusion case $\gamma=1$.  One can see that is not posible to fully match the recovery curve obtained from the CTRW model simply by shifting horizontally the solution for the normal diffusion case, implying that a simple substitution of the form $D\to D' t^{\gamma-1}$ in Eq.~\eqref{eq:Cttheory2} cannot be used to reproduce the behavior of this quantity in the CTRW case. For example, for values of $C_L$ larger than 1/2, say, it would be possible to fit the anomalous diffusion curves reasonably well by a proper shift of the normal diffusion curve. However, this would worsen the agreement for $C_L<1/2$, i.e., the regime corresponding to relatively short times.  One can nevertheless see from the figure that the overall agreement is not too bad. This could provide an empirical justification for a heuristic ``fitting'' procedure of recovery curves based on sBm, especially for experimental comb-like systems characterized by an effective value of $\gamma$ close to 1.

\section{Summary and outlook}
In the present work we obtained an explicit analytical expression for the diffusion coefficient of a particle moving on a comb with randomly varying tooth lengths drawn from a power-law distribution (random comb model). This was done by exploiting the well-known correspondence of the comb model with the CTRW model, whereby the waiting time of the CTRWer was set identical with the time needed by a particle diffusing on a tooth to reach the intersection with the backbone and then move along it. The influence of binding/unbinding processes on the diffusion coefficient was also studied, and a scaling law in terms of the ratio of rate constants for both processes was found. Transport properties are directly related to the specific geometry of the substrate and to the strength and persistence of the binding interactions and could thus provide relevant information about these properties.  Hence, we expect the above results to be useful for quantitative studies of diffusional transport in comb-like systems such as spiny dendrites. In order to mimic these systems in a more realistic way, one could incorporate further sources of spatial disorder into the system, e.g., spatial fluctuations in the separation distance between consecutive teeth. Work in this direction is underway.

Despite the approximations implied by the CTRW model, our analytic results for the diffusion coefficient are in remarkable agreement with Monte Carlo simulations. The analytic expression for the diffusion coefficient was subsequently used to study the behavior of relaxation curves for FRAP processes implemented on the random comb. The agreement was also excellent in this case in spite of the complete absence of free parameters.  The situation studied corresponds to the case in which all the particles were initially placed on the backbone and then some of them were photobleached, as opposed to typical experiments in spiny dendrites, where some particles are still found in the spines after photobleaching. However, no matters of principle prevent one from implementing our initial condition in real experiments.

We also characterized the delay introduced by binding/unbinding processes in terms of a scaling law involving both the ratio of rate constants $\kappa$ and the exponent characterizing the statistical properties of the random comb geometry $\alpha$. Binding/unbinding effects have no influence on the diffusion exponent, but they do change the diffusion coefficient. Similarly, changes in $\alpha$ only have an influence on the diffusion coefficient. In contrast, for $0<\alpha<1$  small changes in $\alpha$ influence both the diffusion exponent and the diffusion coefficient.

As a result of the above, we conclude that the effect of binding/unbinding processes can be mimicked by a change in the comb geometry only when $\alpha>1$. To this end, one needs to find a proper value of the decay exponent $\alpha'$ so that $D_0(\alpha')=D_\kappa(\alpha)$, where both $\alpha,\alpha'>1$. In contrast, when $0<\alpha<1$, this is no longer possible, since the diffusion exponent is also affected, i.e., $\gamma'=(1+\alpha')/2\neq (1+\alpha)/2=\gamma$. In this regime, changes in the geometry have a more profound effect than changes in the binding properties of the system. In the first case, the long time dependence of the MSD is affected, and this should be clearly distinguishable in experiments.    

Finally, we saw that one should be careful when using results based on standard diffusion equations to deal with anomalous diffusion problems. In our case, it was not possible to generate FRAP dynamics on a random comb as given by the CTRW approach by simply replacing the diffusion coefficient with a time-dependent one in the standard diffusion equation, albeit differences appeared to be small in general.  This emphasizes the need of dealing with anomalous diffusion problems by means of bottom-up approaches relying on a solid basis at a microscopic level of description. In this context, our CTRW approach can be straightforwardly extrapolated to study other problems with different initial conditions, boundary conditions, and dimensionality.

\section*{Acknowledgements}
We thank Prof. J. Wlodarczyk for making the raw data of Fig.~4 in Ref.~\cite{Ruszczycki2012} available to us. This work was partially funded by MINECO (Spain) through Grants No. FIS2013-42840-P (partially financed by FEDER funds) (S. B. Y. and E. A.), and by the Junta de Extremadura through Grant No. GR15104 (S. B. Y. and E. A.).  One of us (A.B.) would like to acknowledge the hospitality of the Departamento de F\'{\i}sica at the  Universidad de Extremadura in Badajoz (Spain), where this study was accomplished.

\section*{Appendix A: First passage probability for arbitrary $\theta$ }
\label{Appendix_psin}
In our route to the analytic expression for the first-passage probability $\psi_n(\ell, \theta)$ introduced in Sec.~\ref{subs:alphapos}, we shall invoke some results obtained in the framework of the continuous-time formalism developed in Ref.~\cite{Balakrishnan1995}. We shall formulate the problem in continuous time by first introducing the probability densities $\psi(t)\equiv \psi(t,\ell)$ and $U(t)\equiv U(t,\ell)$. The passage to a formulation in discrete time can be performed via the relations $\lim_{\epsilon\to 0}\int_{n-\epsilon}^{n+\epsilon} \psi(t) dt =\psi_{n}(\ell)$ and $\lim_{\epsilon\to 0}\int_{n-\epsilon}^{n+\epsilon} U(t) dt =U_{n}(\ell)$.

Let us denote by $\varphi(t)$ the pdf for the waiting time between consecutive jumps. In the main text we consider the case of jumps taking place at regular time intervals [$\varphi(t)=\delta(t-1)$)] but the results given below remain valid for an arbitrary pdf provided that its first moment be finite.  A walker placed on the backbone, at $(x=0,y=0)$, say, can reach either the left or the right nearest neighbor site $(x=\pm 1,y=0)$ in many different ways: either
directly by means of a single step (an event which takes place with probability $1-\theta$); or else the walker first jumps upwards to site $(x=0, y=1)$ with probability $\theta$, then returns to $(x=0, y=0)$ after $n-2$ steps (with probability $U_{n-2}$), and finally jumps to $(x=\pm 1,y=0)$ with probability $1-\theta$; or else it first jumps to site $(x=0, y=1)$ with probability $\theta$, then returns to site $(x=0, y=0)$ after $n_1$ steps, then jumps back again to site $(x=0, y=1)$ , then returns to site $(x=0, y=0)$ after $n-3-n_1$ steps, and finally performs the transition to a nearest neighbor site  $(x=\pm 1,y=0)$; and so on. A detailed bookkeeping of all these possibilities leads to the following equation \cite{Balakrishnan1995}:
\begin{align}
\psi(t)=&(1-\theta) \varphi(t)+\int_0^t dt_2\int_0^{t_2} dt_1 \theta \,\varphi(t_1) U(t_2-t_1)(1-\theta)\,\varphi(t-t_2) \nonumber \\
&+ \int_0^t dt_4\int_0^{t_4} dt_3\int_0^{t_3} dt_2 \int_0^{t_2} dt_1\theta \varphi(t_1) U(t_2-t_1) \theta \varphi(t_3-t_2)U(t_4-t_3)   \nonumber \\
&\times\,(1-\theta) \varphi(t-t_4)+\cdots
\end{align}
The multiple convolution structure of the terms on the right hand side suggests that switching to Laplace space may be a convenient strategy; indeed, the Laplace transform of the above equation takes a remarkably simple form, namely, $\tilde \psi(s)= (1-\theta)\tilde\varphi(s)/[1-\theta \tilde\varphi(s) \tilde U(s)]$. Correspondingly, the Laplace transform of the so-called survival probability  $\Psi(t)=\int_t^\infty \psi(\tau) d\tau$, i.e.,  the probability that the waiting time between two consecutive jumps along the backbone be longer than $t$ takes the form 
\begin{equation}
\label{Psis}
\tilde \Psi(s)=\frac{1}{s}\, \frac{\theta\tilde\varphi(s) [1-\tilde U(s)]}{1-\theta \tilde\varphi(s) \tilde U(s)}.
\end{equation}
Using the fact that  $\tilde U(s)=\cosh(\ell+1/2)\xi_0/\cosh(\ell+3/2)\xi_0$ with $\cosh\xi_0=1/\tilde\varphi(s)$ \cite{Balakrishnan1995}, and taking into account that $\xi_0={\cal O}(s^{1/2})$, Eq.~\eqref{Psis} yields the following asymptotic behavior:
\begin{equation}
\tilde \Psi(s)\sim\frac{\theta}{1-\theta} \, \frac{\xi_0 \sinh\ell\xi_0}{2s\cosh\ell\xi_0} \, , \quad s\to 0.
\end{equation}
One thus sees that, for a given $\theta\neq 1/2$,  the probability $\Psi(t)$ is $\theta/(1-\theta)$ times its counterpart for the case $\theta=1/2$.  The same is true for the pdf $\psi(t)$ because of the relation $\Psi(t)=\int_t^\infty\psi(\tau)\,d\tau$, whence Eq.~\eqref{psintheta}
for the corresponding discrete-time probabilities follows.

\section*{Appendix B: Solution of the FRAP problem in Laplace space}
\label{AppendixB}
Since the initial condition \eqref{initcond} is symmetric with respect to the origin, the imposed periodic boundary conditions are equivalent to zero-flux boundary conditions, $j(x \!=\! \pm M/2,t)=0$, whereby the flux of the mobile species (whose divergence is given by the time derivative of the concentration) takes the form
\beq
j(x,t)=-K_\gamma \, ~ _{0}D_t^{1-\gamma}
\frac{\partial c(x,t)}{\partial x} .
\eeq
Hence, zero-flux boundary conditions imply $\left.\frac{\partial c(x,t)}{\partial x}\right|_{x=\pm M/2}=0$.
Given the symmetry of the problem with respect to $x=0$, one has $c(x,t)=c(-x,t)$, and then it is simpler to solve the equivalent problem in the half interval $0\leq x\leq M/2$, whereby the zero-flux boundary condition at $x=-M/2$ is replaced with the same boundary condition at $x=0$, i.e.,
\beq
\left.\frac{\partial c(x,t)}{\partial x}\right|_{x=0}=0 \quad \mbox{ and }\quad \left.\frac{\partial c(x,t)}{\partial x}\right|_{x=M/2}=0.
\eeq
We now introduce the auxiliary quantity $u(x,t)=c_0-c(x,t)$. The Laplace transform $\widetilde{u}(x,s)$ obeys the equation
\beq
\label{bvpEq1}
\frac{d^2 \widetilde{u}}{dx^2}-q_\gamma^2\widetilde{u}=
-\frac{u_0(x)}{K_\gamma s^{1-\gamma}},
\eeq
where $q_\gamma =\sqrt{s^\gamma/K_\gamma}$. In our specific case the initial condition $u(x,0)\equiv u_0$ is
\beq
u_0=
\begin{cases}
c_0, & x \le L/2,\\
0, &  L/2 < x \le M/2,
\end{cases}
\eeq
whereas the boundary conditions are
\beq
\label{bvpEq2}
\left.\frac{\partial \widetilde{u}}{\partial x}\right|_{x=0}=0 \quad \mbox{ and }\quad \left.\frac{\partial \widetilde{u}}{\partial x}\right|_{x=M/2}=0.
\eeq
Let us now rescale the length variables with $q_\gamma$, that is, we define $\hat{x}\equiv q_\gamma x$, $\hat{L} \equiv q_\gamma L$ and $\hat{M} \equiv q_\gamma M$. Further, let us introduce $\hat{u}_0(\hat{x})=u_0( \hat{x}/q_\gamma)$; the problem described by Eqs. \eqref{bvpEq1}-\eqref{bvpEq2} can then be written as follows:
\beq
\frac{d^2 \widetilde{u}(\hat{x},s)}{d \hat{x}^2}
-\widetilde{u}(\hat{x},s)=- \frac{\hat{u}_0(\hat{x})}{s},
\eeq
with
\beq
\hat{u}_0(\hat{x})=
\begin{cases}
c_0, & \hat{x} \le \hat{L}/2,\\
0, &  \hat{L}/2 < \hat{x} \le \hat{M}/2,
\end{cases}
\eeq
and
\beq
\left.\frac{\partial \widetilde{u}}{\partial \hat{x}}\right|_{\hat{x}=0}=0 \quad \mbox{ and }\quad \left.\frac{\partial \widetilde{u}}{\partial \hat{x}}\right|_{\hat{x}=\hat{M}/2}=0.
\eeq
The Green's function $G(\hat{x},\eta;s)$ for the above Sturm-Liouville problem fulfils the equation
\beq
\frac{\partial^2 G(\hat{x},\eta;s)}{\partial \hat{x}^2}
-G(\hat{x},\eta;s)= \delta(\hat{x}-\eta),
\eeq
as well as the requirements
\begin{align}
& G(\hat{x}\to \eta^-,\eta;s)=G(\hat{x}\to \eta^+,\eta;s), \\
  & \left. \frac{\partial G(\hat{x},\eta;s)}{\partial \hat{x}}
  \right|_{\hat{x}\to \eta^-}-
  \left. \frac{\partial G(\hat{x},\eta;s)}{\partial \hat{x}}
  \right|_{\hat{x}\to \eta^+}=1.
\end{align}
The solution of the above problem can be computed by standard techniques \cite{Arfken2013ch10}. The final result is
\beq
G(\hat{x},\eta;s)=
\begin{cases}
  & \frac{1}{2} (e^{\eta}+e^{\hat{M}-\eta})
  \left(e^{\hat{x}}+e^{-\hat{x}}\right)/(1-e^{\hat{M}}), \quad \hat{x} \le \eta,\\
  & \frac{1}{2} (e^{\eta}+e^{-\eta})
  \left(e^{\hat{x}}+e^{\hat{M}-\hat{x}}\right)/(1-e^{\hat{M}}), \quad \hat{x}\ge \eta.\\
\end{cases}
\eeq
The solution of the original problem in terms of rescaled variable can then be expressed as follows
\beq
\widetilde{u}(\hat{x},s)=-\frac{1}{s}\int_0^{\hat{M}/2}
G(\hat{x},\eta;s) \hat{u}_0(\eta)\, d\eta.
\eeq
Performing the integration and undoing the length rescaling, one is finally left
with Eq.~\eqref{f-conc}.

%

\end{document}